\pdfoutput=1

\documentclass[twocolumn]{aastex6}

\usepackage{float}
\usepackage{appendix}
\usepackage{chngcntr}
\usepackage{amsmath}
\usepackage{color}
\usepackage{subfigure}
\usepackage{graphicx}

\newcommand{\cn}[1]{{{\color{red} (cite(s) needed)}}}
\newcommand{\fc}[1]{{\emph{(add cite to bibliography)}}}
\newcommand{\KpIp}[1]{{Kirkpatrick \emph{(in progress)}}}
\newcommand{\nd}[1]{{\emph{***Not done...}}}
\newcommand{\agnf}[1]{$L_{\rm{AGN}} / L_{\rm{IR}}$}
\newcommand{\emit}[1]{\rm{emit}}
\newcommand{\obs}[1]{\rm{obs}}
\newcommand{\data}[1]{\rm{data}}
\newcommand{\model}[1]{\rm{model}}
\newcommand{\um}{\,$\mu$m}

\begin{document}
\title{Simulations Find Our Accounting of Dust-Obscured Star Formation May Be Incomplete}
\author{Eric Roebuck\altaffilmark{1}, Anna Sajina\altaffilmark{1}, 
Christopher C. Hayward\altaffilmark{2}, Nicholas Martis\altaffilmark{1},
 Danilo Marchesini\altaffilmark{1},Nicholas Krefting\altaffilmark{1}, Alexandra Pope\altaffilmark{3}}
\altaffiltext{1}{Department of Physics and Astronomy, Tufts University, Medford, MA 02155, USA; eric.roebuck@tufts.edu} 
\altaffiltext{2}{Center for Computational Astronomy, Flatiron Institute, 162 5th Avenue, New York, NY 10010, USA} 
\altaffiltext{3}{Department of Astronomy, University of Massachusetts Amherst, Amherst, MA, USA} 

\begin{abstract}
The bulk of the star-formation rate density peak at cosmic noon was obscured by dust. How accurately we can assess the role of dust obscured star-formation is affected by inherent biases in our empirical methods -- both those that rely on direct dust emission and those that rely on the inferred dust attenuation of starlight. 
We use a library of hydrodynamic simulations with radiative transfer to explore these biases.
We find that for IR luminous galaxies that are in rapidly quenching systems (e.g. post-coalescence) standard luminosity-to-SFR relations can strongly overestimate the true SFRs. 
We propose using the $L_{IR}/L_{1.6}$ color to both help identify such systems and provide more accurate SFRs. 
Conversely, we find that the diagnostic UVJ plot misidentifies a subset of dusty star-forming galaxies.
This is due to variability in the effective attenuation curves including being much grayer in the optical-to-near-IR regime than the Calzetti starburst law. This is in agreement with recent observations of IR-selected galaxies at cosmic noon. 
Our results support the view that we need a panchromatic approach from the rest-frame UV through the IR and SED modeling that includes realistic SFHs and allows for variable attenuation curves if we want to fully account for dust obscured star-formation across the epochs of greatest galaxy build-up.
\end{abstract}

\section{Introduction}
Extragalactic studies rely on our ability to convert observables into physical parameters of galaxies. A fundamental such parameter is the star-formation rate (SFR).
Indeed, a core result in extragalactic astrophysics is that the star-formation rate density of the universe peaked at $z\sim$1-3 and has since decreased by $\sim$10 fold \citep[see][for a review]{MadauDickinson2014}. This peak is known as ``cosmic noon" and it is the epoch during which the bulk of stellar and black hole mass were build-up. However, the precision studies that we need in order to better understand this build-up and the subsequent decline in overall activity, are hindered by the systematic uncertainties in estimating SFRs. These uncertainties can be as high as an order of magnitude \citep[see e.g.][]{Muzzin2009}. 

One systematic affecting the derivation of a galaxy's SFR is its dependence on the unknown star-formation history (SFH). In systems such as interacting galaxies, the SFH can be far from the simple parametrizations typically used in SED fitting codes \citep[e.g.][]{Boquien2014,Sklias2017}. Another systematic is that estimating the SFR requires accounting for both dust obscured and unobscured star-formation, which requires knowledge of the dust attenuation curve. The later depends on the unknown beyond the local Universe dust properties \citep[see e.g.][]{Sajina2009} as well as the relative star-dust geometry \citep[e.g.][]{WittGordon1996,Calzetti2000,Draine2001,Seon2016}. Over the last few years, progress has been made on both the issue of accounting for the role of the star-formation history and a potentially variable attenuation curve. For example, \citet{Boquien2016} have shown promising results with adaptive composite conversion relations to derive SFRs from observed FUV and IR luminosities. Their relations aim to account for the role of older stellar populations in dust heating. New SED fitting codes allow for more flexible SFH modelling \citep{Leja2017} which helps us understand the biases inherent in using simple parametrizations thereof. Commonly used codes such as MAGPHYS and CIGALE \citep{daCunha2008,Noll2009} treat attenuation toward the birth-clouds and in the diffuse ISM separately, which allows for more flexible effective attenuation curves. However, recent work suggests there may need to be additional modification to allow for greater flexibility \citep{CharlotFall2000,LoFaro2017,Buat2018}. Both observational and theoretical studies have shown that such flexibility is required since a single attenuation curve is not appropriate for all star-forming galaxies either locally or at high-$z$ \citep[e.g.][]{Kriek2013,Chevallard2013,Battisti2017a,LoFaro2017,Salim2018,Narayanan2018,Buat2018}. 

Hydrodynamic simulations of galaxies with radiative transfer treatment allow us to test the conversion of observables into physical properties such as SFRs. Here the star-formation histories are exactly known, as are the intrinsic dust properites. This is highlighted in recent work on relating the observable IRX-$\beta$ relation \citep{Safarzadeh2017} as well as the $L_{IR}$ vs. instantaneous and 100\,Myr average SFR in such simulations \citep{Hayward2014}. In previous papers we have used such simulations to explore how well observable diagnostics recover the true AGN power of galaxies \citep{Snyder2013,Roebuck2016}. In this paper, we examine how well observable UV through IR diagnostics and SED fitting methods recover the true SFRs of our simulations. We focus on massive, gas-rich systems both evolving in isolation and involved in major mergers. These give us a variety of star-formation histories and overall dust levels. Our simulated galaxies approximate massive galaxies at cosmic noon and beyond which were typically more gas-rich both intrinsically and due to enhanced gas accretion \citep[e.g.][]{Tacconi2010,Yan2010,Scoville2017}. 

Our paper is organized as follows.
In Section\,\ref{sec:simulations} we present our simulations. In Section\,\ref{sec:results} we examine how the true SFRs of our simulated galaxies match up with common observable diagnostics. We present a new method for correcting the IR luminosity for the effective age of the stars heating the dust using the observed $L_{IR}/L_{1.6}$ colors. 
We show that the rest-frame U-V vs. V-J colors are not very effective at probing the evolutionary stage of our simulated galaxies due to the effects of variable attenuation curves that commonly are greyer than Calzetti curves redward of the $V$-band. In Section\,\ref{sec:discussion}, we use a toy model to further explore the observed trends in effective attenuation curves. In this Section, we also compare our results with recent observational results and discuss the caveats in our analysis. Finally, in Section\,\ref{sec:conclusions} we summarize our results and conclusions.  

\section{Simulations} \label{sec:simulations}
The model library used in this analysis was generated using the \textsc{gadget-2} cosmological $N$-body/smoothed particle hydrodynamics (SPH) code \citep{Springel2005gadget}. These models are post-processed using the \textsc{sunrise} \citep{Jonsson2006, Jonsson2010} radiative transfer suite to generate panchromatic spectral energy distributions (SEDs) at 10-100 Myr increments and for 7 isotropic viewing perspectives. The simulations used in this paper were presented in previous works (see Table~\ref{tab:modeltable}).

\subsection{Hydrodynamic Simulations} \label{sec:hydro}

\textsc{gadget-2} \citep{Springel2001} uses a fully conservative \citep{Springel2002} modified TreeSPH \citep{Hernquist1989} hydrodynamics, and accounts for radiative heating and cooling \citep{Katz1996}. Star-formation depends on the volume-density adjusted Kennicutt-Schmidt (KS) relation $\rho_{\rm SFR} \sim \rho_{\rm gas}^{1.5}$ \citep{Kennicutt1998a}, normalized to produce the galaxy scale KS relation. We use the two-phase subresolution ISM model of \citep{Springel2003}, which accounts for supernovae feedback \citep{Cox2006b}. Gas particles self-enrich (closed-box) with a yield of $y \sim 0.02$ according to the star formation rate. The distribution of age and metalicity of the star and gas particles is calibrated to observations \citep{Rocha2008,Jonsson2010}.

Super massive black hole (SMBH) accretion and feedback are done using the model of \citet{Springel2005feedback}. Black hole sink particles begin with initial mass $10^5 \ \rm M_{\odot}$. They accrete at the Eddington-limited Bondi-Hoyle rate, where 5\% of the luminous energy is returned to the ISM as thermal feedback calibrated to the $M$-$\sigma$ \citep{DiMatteo2005}. The radiative efficiency is 10\% ($L_{\rm bol}=0.1 \dot{M}_{\rm BH} c^2$). Additional information on \textsc{gadget} can be found in \citet{Springel2005gadget}.

\subsection{Radiative Transfer}
We perform radiative transfer using the 3D Monte Carlo code \textsc{sunrise} \citep{Jonsson2006}. Star particles are treated as single-age stellar populations. Those present in the initial conditions are assigned ages and metallicites as detailed in the works in which the simulations were originally presented (see Table \ref{tab:modeltable}). We have checked that SEDs are insensitive to these reasonable variations in the assumed star formation history of the initial stellar populations because for single-aged stellar populations with age $>>100$ Myr, the SED varies slowly with time. Star particles aged $>10 \ \rm{Myr}$ are assigned \textsc{starburst99} \citep{Leitherer1999} template SEDs, whereas those with ages $<10 \ \rm{Myr}$ are assigned {\sc mappingsiii} templates from \citet{Groves2008} that include H\textsc{II} and photodissociation regions (PDRs). For a detailed description of the {\sc mappingsiii} implementation within {\sc sunrise}, we refer the reader to \citet{Jonsson2010}. Black holes are assigned luminosity-dependent AGN SED templates of \citet{Hopkins2007}, which are empirical templates based on observations of unreddened quasars \citep{Richards2006b}.

Once the source (star and AGN) particles are set the \textsc{gadget} gas-phase metal density is projected onto a 3D octree grid initially at 200 kpc on a side. The dust density assumes 40\% of metals are in the form of dust \citep{Dwek1998}, and the initial mass fraction in metals is Z=0.01 \citep{Snyder2013}.\footnote{Because we assume a constant-dust-to-metal ratio, gas particles contain dust regardless of their temperature (in contrast to, e.g., \citet{Cochrane2019}, who assume that gas cells with $T > 10^6$ K contain no dust due to grain sputtering). Imposing such a cut would make little difference because in these idealized simulations without a hot halo or stellar feedback-driven winds, a small fraction of the metals (and thus dust) reside in hot gas, and due to this dust typically being located at large radii, it contributes little to the overall column density along a given line of sight. We prefer to not utilize a temperature cut for simplicity, and because these simulations do not resolve the phase structure of the ISM by construction. In reality, gas particles with $T > 10^6$ K would contain cold ‘subclumps’ that may contain dust, so assuming that such particles have no dust due to sputtering would also not be completely correct.}
We assume the Milky Way (MW) dust model of \citet{Draine2007}.
As detailed in the works in which the simulations were previously presented, the grid refinement parameters were chosen such that the SEDs were converged to within $\sim$10\%; see \citet{Jonsson2010} for more details about the grid refinement scheme and the works listed in Table \ref{tab:modeltable} for the specific grid refinement parameters employed.

From the source particles $10^7$ panchromatic photon packets are propagated through the grid accounting for absorption and scattering by dust. The radiation absorbed is re-emitted in the infrared (IR) assuming large grains are in equilibrium. Half of the small grains are assumed to emit thermally, the rest emit as PAHs using \citet{Groves2008} template. This fraction is calibrated \citep{Jonsson2010} to match mid-IR flux ratios from SINGS \citep{Dale2007}. The dust re-emission is also propagated through the grid to allow for self-absorption. The output of the simulation are UV-mm SEDs at 7 isotropic viewing angles at 10-100 Myr increments. 

The viewing angles are isotropically distributed to cover the 4$\pi$ around each simulated galaxy. We note that this means we do not necessarily have specific ``edge-on" and ``face-on" views for our isolated disk galaxies which is known to affect the dust attenuation (and hence colors) in disk galaxies \citep[e.g.][]{Wang2018}. We have explored the role of viewing perspective for one of our mergers in \citet{Snyder2013} and found that the effect is essentially a variation in effective $A_V$
-- some lines-of-sight are dustier than others. Across the star-formation histories of our simulated galaxies, the effective $A_V$ also changes significantly. Indeed, we found that variation between viewing angles throughout this paper were smaller than the trends driven by the star-formation history. Therefore for clarity, we chose to focus on the quantities averaged over the 7 isotropic viewing angles such that we can concentrate on the trends with star-formation history rather than those with viewing perspective. 

\begin{deluxetable}{cccc}
\tabletypesize{\footnotesize}
\tablecolumns{4} 
\tablewidth{0.45\textwidth}
 \tablecaption{Initial simulations setup \label{tab:modeltable}}
 \tablehead{
 \colhead{Model} & \colhead{$\log(M_*/{\rm M_{\odot}})$} & \colhead{$f_{\rm gas,init}$} & \colhead{References\tablenotemark{a}}} 
 \startdata 
 Isolated M5 & 10.6 & 0.6 & c5[H13], M5[R16] \\ 
 Isolated M6 & 11.2 & 0.6 & Iso. Disc[H11], M6[R16] \\
 Isolated M7 & 10.3 & 0.8 & b5[H12,H13], M7[R16] \\
 Isolated M8 & 10.9 & 0.8 & b6[H12,H13], M8[R16] \\
 Merger M5e & 10.9 & 0.6 & M5\tablenotemark{a,b}[R16] \\
 Merger M6e & 11.5 & 0.6 & Merger[H11], M6\tablenotemark{a,b}[R16] \\
            &  & & Highly Obscured[S13]
 \enddata
 \vspace{0.01in}
\tablenotetext{a}{References to prior publications using these models with associated \\
labels (note that in R16 we tabulated the isolated disks, where \\
they served as progenitors for equal mass mergers; e.g. M5e is the \\
equal mass merger of two ``M5" discs in R16). The references are: \\
H11 \citep{Hayward2011}; H12 \citep{Hayward2012}; \\
H13 \citep{Hayward2013}; S13 \citep{Snyder2013}; \\
H15 \citep{Hayward2015}; and R16 \citep{Roebuck2016}.}
 \tablenotetext{b}{The initial stellar masses for the mergers are the sum of the two \\
 progenitors using the `e' orbit of \citep{Cox2006a}. }
\end{deluxetable}

\subsubsection{ISM Treatment} \label{sec:ismtreatment}
The ISM as described in Section~\ref{sec:hydro} uses the two-phase model of \citet{Springel2003}, where each element contains a warm ($>10^5$ K) and cold ($<10^4$ K) gas component. Only a single density and ``effective pressure" is evolved. Within {\sc sunrise}, sub-grid dust clumpiness is treated using one of two extreme assumptions. One is to assume the cold dense clouds implicit in the \citet{Springel2003} model have negligible volume filling factor for the purposes of the radiative transfer calculations. This assumption is hereafter referred to as multiphase-on and denoted {\sc mp-on}. The other extreme is to consider that the cold clouds have 100\% volume filling factor, so the total metal mass from {\sc gadget-2} is used when computing the dust mass (multiphase-off, denoted {\sc mp-off}). In both cases, the dust density is assumed to be uniform on the scale of the SPH smoothing length.

\begin{figure}[!h]
\includegraphics[width=0.45\textwidth]{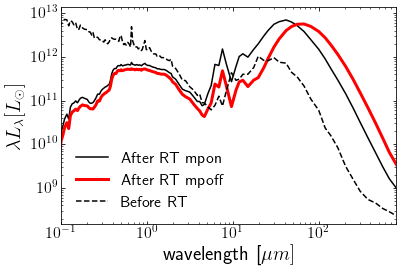}
\caption{The effect of the \textsc{sunrise} radiative transfer (RT) calculation for both the {\sc mp-on} and {\sc mp-off} ISM treatments (see text for detail). The dashed SED is the stellar input SED before dust attenuation. These are all shown for the peak dust attenuaton epoch of the M6e simulation. The {\sc mp-off} case has overall higher $A_V$ but not significantly different attenuation curve. It also shows an infrared SED that is shifted toward colder temperatures, but the overall IR luminosity is not significantly different. }
\label{fig:SED_summary}
\end{figure}

Figure\,\ref{fig:SED_summary} illustrates the effect of the different treatments on the output from the radiative transfer within \textsc{sunrise}. We see that the {\sc mp-off} version leads to somewhat higher $A_V$ as expected as it is the less clumpy configuration (i.e. closer to a dust screen). This configuration also leads to overall colder effective dust temperatures. In this work we cannot fully explore the effects of the two treatments given that we only have both versions run for the simulation shown in Figure\,\ref{fig:SED_summary}. We note however that despite the above differences, we found that substituting the {\sc mp-on} version for {\sc mp-off} for this simulations in any of the subsequent plots has a negligible effect on our conclusions. This is because the differences between the two version in overall IR luminosity or the shape of the attenuation curve at various timesteps are much smaller than the broader trends we are finding in this paper. 

\subsubsection{AGN Treatment} \label{sec:AGNtreat}

The default treatment of the AGN luminosity (AGN1x) assumes the black hole accretion rate from \textsc{gadget-2} (Section~\ref{sec:hydro}). For the purposes of isolating the stellar contribution to the SED in this analysis, we can artificially suppress the AGN luminosity (AGN0x). Thermal feedback from the AGN (radiative efficiency 10\%, Section~\ref{sec:hydro}) is always included whether or not the AGN luminosity is engaged.

\begin{figure*}[!h]
\includegraphics[width=0.45\textwidth]{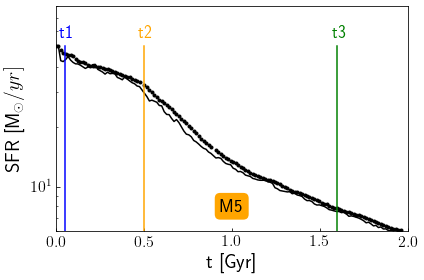}
\includegraphics[width=0.45\textwidth]{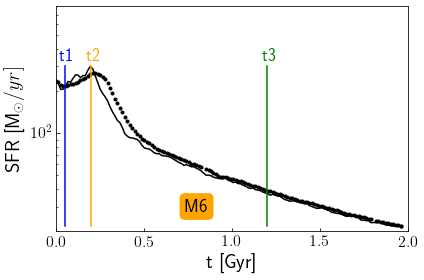}\\
\includegraphics[width=0.45\textwidth]{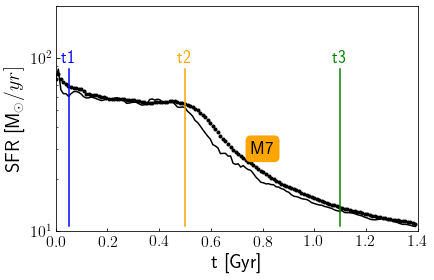}
\includegraphics[width=0.45\textwidth]{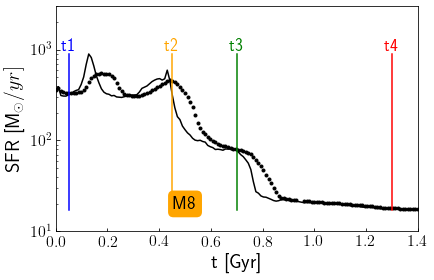}\\
\includegraphics[width=0.45\textwidth]{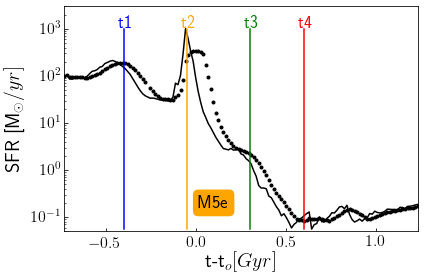}
\includegraphics[width=0.45\textwidth]{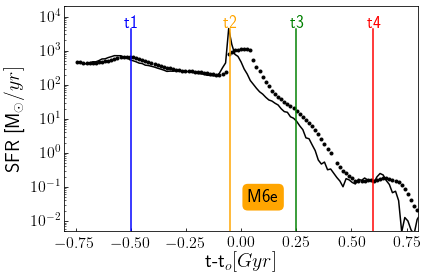}
\caption[Overview of simulations]{Here we highlight the wide range of star formation histories covered by our simulations including both isolated galaxies ({\it top two rows}), and  major mergers ({\it bottom row }).  We show both the instantaneous SFRs (thin solid curves) and those averaged over 100\,Myrs, $\langle SFR \rangle_{100Myr}$ hereafter (thick black dots). The indicated times ($t_1-t_4$) are described in more detail in the text.  }
\label{fig:time_evolution}
\end{figure*}

\subsection{Simulation Library}
Table\,\ref{tab:modeltable} lists the specific models we use as well as their initial stellar mass and gas fractions. We also give references to previous publications where these models are discussed; see those works for full details of the initial conditions. This library is drawn from the compilation in \citet{Roebuck2016}. Given the goals of this paper, we exclude models that are not very dusty (rarely reach $A_V$\,=\,1) or where the SFH sampling is too limited and/or sparse. We emphasize that the requirement that we often reach $A_V>1$ leads to our simulations being gas-rich - i.e. $z\sim3$ analogues \citep[e.g.][]{Hayward2013}. Looking at Table\,\ref{tab:modeltable}, our models are all massive and gas-rich. They are likely to have proxies among massive dusty galaxies (especially at higher redshifts where gas fractions were higher). The converse is not true -- our simulation library probes a limited range in initial conditions and therefore we do not expect all observed massive, dusty galaxies to have proxies among our simulations. 

We also exclude models where the IR luminosity is at times dominated by an AGN \citep[see][for discussion on the role of AGN in dust heating]{Roebuck2016}, since here we focus on the role of dust obscured star-formation. To that effect, for the Merger M6e we use the the runs with the AGN contribution to the SED removed (AGN0x, Section~\ref{sec:AGNtreat}). The AGN was not explicitly removed in the isolated galaxy simulations or the other merger simulation, but in both cases its contribution is negligible as its power output is significantly below that of the stars throughout the UV-IR regime at all times in the simulation. We chose to keep the simulations where the AGN is not explicitly removed since this allows us to probe a wider range of properties than using only models with AGN0$\times$ (we examined other AGN0$\times$ models which were excluded from the final sample as they reach $A_V>1$ only briefly)\footnote{ 
This also mimics real observations, where dominant AGN are easier to identify but sub-dominant ones are not and are likely present to some degree in most dusty star-forming galaxy samples.}.

While building a larger more representative custom library was beyond the scope of this project, the library used here has been shown to reproduce various galaxy observables in previous works (see references in Table\,\ref{tab:modeltable})  and is well-studied. This makes it a natural choice for this pilot study. 
The subset of models used in this analysis are of the {\sc mp-off} ISM treatment as default. We discuss the effects of ISM treatment in Section~\ref{sec:ismtreatment} on M6e for reference. 

\begin{figure*}
\includegraphics[width=0.9\textwidth]{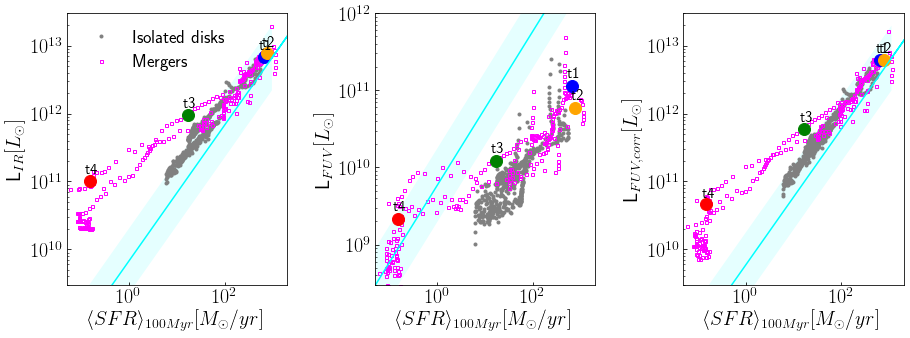}
\caption{Here we compare the simulated 100\,Myr-averaged SFR for each timestep of our simulations with $L_{IR}$, $L_{FUV}$ and the dust-corrected $L_{FUV,corr}$ done following \citet{Boquien2016}. For the M6e merger we also mark the four representative timesteps from Figure\,\ref{fig:time_evolution}. In all panels the cyan lines are the relations from \citet{KennicuttEvans2012} with the cyan bands representing $\pm$3\,$\times$ the SFR. For the isolated disk galaxies both $L_{IR}$ and $L_{FUV,corr}$ are reasonable proxies for $\langle SFR\rangle_{100Myr}$, while $L_{FUV}$ alone is not a good proxy as these simulations have high dust obscuration ($A_V>1$) throughout.  The  mergers follow the same trends for the early stages of the mergers (until coalescence) but strongly deviate from these relations in the rapidly quenching post-coalescence regimes. Here stellar populations on timescales $>100$Myr dominate the dust heating. Note that for such massive, gas-rich mergers this is true even for IR-luminous stages ($L_{IR}\approx10^{11}-10^{12}$\,L$_{\odot}$).}
\label{fig:lum_vs_sfr}
\end{figure*}

\section{Results} \label{sec:results}
\subsection{Star-formation histories \label{sec:sfh}}
Figure\,\ref{fig:time_evolution} shows the star-formation histories (SFH) of our simulations. For the mergers this is given relative to $t_o$, the coalescence time, which is defined as the moment the two black holes merge. In each plot, we highlight a few epochs that are representative for the SFHs and also help indicate the direction of time evolution in subsequent plots. These are as follows:
\begin{itemize}
    \item The first epoch, $t_1$, is representative of the pre-coalescence epoch for the mergers and is placed on top of the first passage SFR bump.  We tested that the qualitative behavior shown throughout this paper is unaffected by moving this point somewhat earlier or later.  For the isolated galaxies, this point is placed in an arbitrary early stage in each simulation.
    \item The second epoch, $t_2$, represents the peak SFR for the mergers (near coalescence). For the isolated galaxies, it represents a bump or a kink in their SFH. For both isolated galaxies and mergers, $t=t_2$ marks  an epoch after which the rate of quenching (decrease of SFR) increases.   
    \item The third epoch, $t_3$, exemplifies this quenching regime where the instantaneous SFR has declined significantly since its peak but is still non-negligible. For the isolated galaxies, this drop is roughly $10\times$, whereas for the mergers where the post-peak quenching is more rapid, this drop is roughly $100\times$.
    \item The fourth epoch, $t_4$, is when the level of ongoing star-formation has reached effectively zero, but we expect dust heating contribution from intermediate age stars. This is a classic post-starburst system. Most isolated galaxy simulations are not run long enough to reach this stage, so lack a $t=t_4$ point. The exception is M8 which while not reaching a negligible SFR, reaches a stage when the rate of quenching has again decreased and the SFH has flattened. Here $t=t_4$ marks this new regime at the end of the simulation.  
\end{itemize}

\subsection{The role of SFH in luminosity-SFR conversion relations \label{sec:lum_vs_sfr}}

Here we compare the simulated $L_{\rm{IR}}$\footnote{Defined as the integral of the SED over 3-1100\,$\mu$m.}, $L_{FUV}$ and  the dust-corrected $L_{FUV,corr}$ with the the 100Myr averaged SFR ($\langle $SFR$\rangle_{\rm 100Myr}$). The values from our simulations are compared with the following relations:
\begin{align}
    \log(SFR/[M_{\odot}/yr])=\log(L_{IR}/[erg/s])-43.41\\
    \log(SFR/[M_{\odot}/yr])=\log(L_{FUV}/[erg/s]-43.35
\end{align}
where in the case of $L_{FUV,corr}$ we substitute $L_{FUV,corr}$ for $L_{FUV}$ above. The dust-corrected FUV luminosity is given by:
\begin{equation}
L_{FUV,corr}=L_{FUV}+k_{IR}L_{IR},
\end{equation}
where $k_{IR}$ is the dust correction factor which we detail further below.  These are the relations from the review article \citet{KennicuttEvans2012} (KE12 hereafter). We refer the reader to that article for further details and references. Some salient details however are that the relations in KE12 assume a Kroupa IMF, same as our simulations, and they assume a constant SFH over the past 100\,Myrs, which is not equivalent to averaging the SFR over the past 100Myrs in our simulations in regimes where the SFH varies rapidly (see Figure\,\ref{fig:time_evolution}). For example, \citet{Boquien2014} find that, in actively star-forming galaxies, both the FUV and IR luminosities are predominantly sensitive to much younger stellar populations (typically $\approx$10\,Myrs).  Using SFRs averaged over the past 100\,Myrs is a convenient quantity to compare to standard relations here allowing us to further explore the role of the SFH in these relations. In particular, if the SFH is rising within the past 100\,Myrs then younger populations will dominate this average, if falling then older populations will dominate it.  Using the 100\,Myr timescale also confines us to the conventional timescales of massive, short lived O/B stars.  If the simulated IR or dust-corrected FUV luminosities are higher than this 100\,Myr averaged SFR, then this signals the contribution of later stellar types to the dust heating. 

Figure\,\ref{fig:lum_vs_sfr} ({\it left}) shows $L_{IR}$ vs. $\langle $SFR$\rangle_{\rm 100Myr}$. During the actively star-forming stages, our simulations tend to fall slightly above the KE12 relation but still within a factor of 3 (shaded region). However, in the rapidly quenching post-coalescence stages of the mergers, significant dust heating by intermediate-age A stars leads to strong offsets in this relation. This heating is well known \citep[e.g.][and references therein]{Hayward2014}. However, our results highlight that this can happen at high IR luminosities (LIRG/ULIRG regime) in massive gas-rich systems as our M5e and M6e simulations.  As seen in Figure\,\ref{fig:time_evolution}, in the quenching regime, the 100Myr-averaged SFR exceeds the instantaneous SFR (or any shorter timescale average) therefore would make the discrepancy here worse. We tested that averaging on longer timescales up to 500Myr at $t=t_4$ is needed for better agreement. Obviously in this regime, the dust is significantly heated by older stellar populations such as A-type stars. 

Figures\,\ref{fig:lum_vs_sfr} ({\it middle}) shows $L_{\rm FUV}$ vs. $\langle $SFR$\rangle_{\rm 100Myr}$. Here we find the reverse trend. During the active star-forming regimes, the observed $L_{\rm FUV}$ strongly underestimates the SFR. However, in the less dusty post-coalescence epochs for the mergers (both mergers drop below $A_V\approx1$ just prior to $t=t_3$), it is a reasonable proxy for the SFR. 

Figure\,\ref{fig:lum_vs_sfr} ({\it right}) shows the dust corrected FUV luminosity vs. $\langle $SFR$\rangle_{\rm 100Myr}$. This correction follows \citet{Boquien2016} who argue that $k_{IR}$ should not be a constant as commonly assumed \citep[see e.g.][and references therein]{KennicuttEvans2012} but is a linear function of the FUV-near-IR colors\footnote{We use their FUV-H color relation. Specifically, $k_{IR}=0.998-0.092(FUV-H)$}. which is a proxy for sSFR. This composite relation gives substantially better estimates of the true SFRs than the unobscured $L_{FUV}$, especially for the isolated galaxies and is comparable to $L_{\rm{IR}}$. However, it again overestimates the SFR for the rapidly quenching, post-coalence stages of the mergers (although performing slightly better here than $L_{IR}$ alone).  

\begin{figure}
\includegraphics[width=0.45\textwidth]{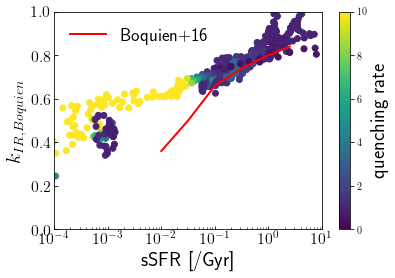}
\includegraphics[width=0.45\textwidth]{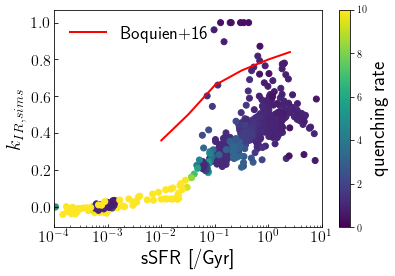}
\caption{{\it Top: } The $k_{IR}$ relation computed following \citet{Boquien2016} and compared with the sSFR for each timestep for each of our simulation color-coded by quenching rate. The red curve is the $k_{IR}$-sSFR relation from \citet{Boquien2016}.  We find that our simulations agree with this relation in low quenching rate regimes (including all isolated disk galaxies), but strongly overestimate it for high quenching rate regimes (post-coalescence in the major merger simualtions). The clump of low quenching rate at low sSFR represents the end stages of each merger simulation. {\it Bottom: } The same plot now showing the $k_{IR}$ needed in the simulations to have $L_{FUV,corr}$ correctly predict the 100Myr-averaged SFR. Note the roughly factor of two offset for the low quenching rate, isolated galaxies compared to \citet{Boquien2016} and the drastically lower $k_{IR}$ in the post-coalescence, high quenching rate regime. 
\label{fig:kIR}}
\end{figure}

\subsubsection{$L_{FUV,corr}$ using $L_{IR}/L_{1.6}$}

Since the relations from \citet{Boquien2016} are meant to correct for SFH, it is surprising that we see this offset in the post-coalescence stages. We believe this is the result of the sample used in \citet{Boquien2016} being isolated star-forming disks and lacking the extremely rapidly quenching post-coalescence stages of our major merger simulations.  To test this theory, in Figure\,\ref{fig:kIR} {\it left}, we compared the $k_{IR}$ relation computed following \citet{Boquien2016} with the sSFR (computed using the 100Myr-averaged SFR) for each timestep for each of our simulation color-coded by quenching rate. This quenching rate is defined as the ratio of the 500Myr-averaged SFR to the 100Myr-averaged SFR.  Our simulations compare well with the expected relation from \citet{Boquien2016},  in low quenching rate regimes (including all isolated disk galaxies), but strongly overestimate $k_{IR}$ for high quenching rate regimes (post-coalescence in the major merger simulations). 
In Figure\,\ref{fig:kIR} {\it right} we show the same plot but now with the $k_{IR}$ needed to have $L_{FUV,corr}$ match the 100\,Myr-averaged SFR in Figure\,\ref{fig:lum_vs_sfr}. We note that we have roughly a factor of two offset between this value and the Boquien-corrected value. This likely has to do with the precise sample used here vs. the sample in Boquien et al, their level of dustiness and the details of their respective SFHs. We agree better in this regime with the average value of $\langle k_{IR}\rangle =0.46\pm 0.12$ derived by \citet{Hao2011}.
As expected, in the high quenching rate, post-coalescence regime the required $k_{IR,sims}$ are indeed much lower than the corresponding $k_{IR,Boquien}$, which were not calibrated for this regime.

\begin{figure}
\centering
\includegraphics[width=0.5\textwidth]{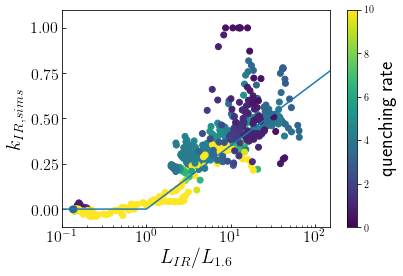}
\caption{The $L_{IR}/L_{1.6}$ ratio is a good predictor of $k_{IR,sim}$ with a relation described in the text. 
\label{fig:irratio_corr}}
\end{figure}

The $k_{IR,sims}$ in Figure\,\ref{fig:kIR} {\it right} were derived with prior knowledge of the SFR from the simulations. We need an observable-based means of deriving this value. In Figure\,\ref{fig:irratio_corr} we show that we have a strong $L_{IR}/L_{1.6}$ correlation with $k_{IR,sims}$. In particular we find that  $k_{IR}=\log(L_{IR}/L_{1.6})*0.35$ where any value below 0 is re-set to zero, and any value above 1 is re-set to 1.0. 

\begin{figure*}
\includegraphics[width=0.3\textwidth]{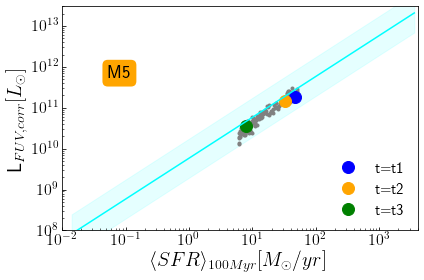} 
\includegraphics[width=0.3\textwidth]{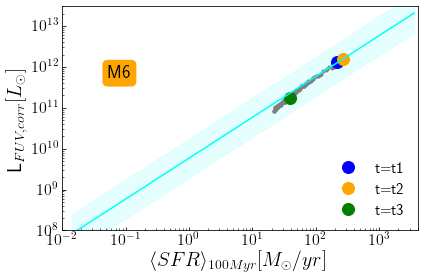}  
\includegraphics[width=0.3\textwidth]{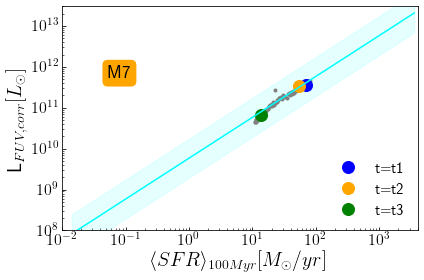}  \\
\includegraphics[width=0.3\textwidth]{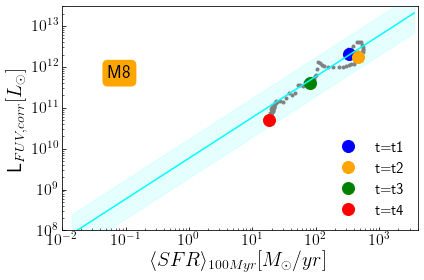}  
\includegraphics[width=0.3\textwidth]{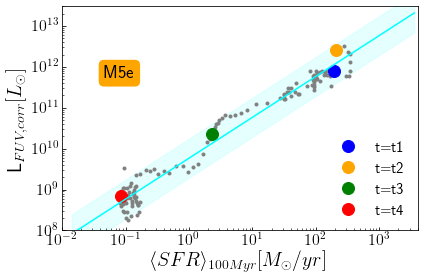}  
\includegraphics[width=0.3\textwidth]{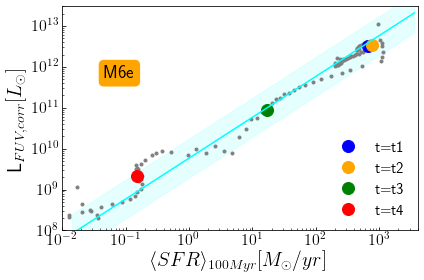} 
\caption[Empirical relation corrected for SFH]{Here $L_{FUV,corr}=L_{FUV,obs}+k_{IR}L_{IR}$ where $k_{IR}$ is derived from the $L_{IR}/L_{1.6}$ ratio as discussed in the text. With this correction we arrive at good agreement between the simulated $\langle$SFR$\rangle_{100Myr}$ and the $L_{FUV,corr}$ luminosities. This is true for both isolated disks and major merger simulations including in the rapidly quenching post-coalescence regimes. The thick cyan line is the conversion relation from KE12 with the shaded region representing $\pm$3$\times$ the derived SFR.
\label{fig:lircorr}}
\end{figure*}

In Figure\,\ref{fig:lircorr} we set $L_{\rm FUV,corr}=L_{\rm FUV,obs}+k_{IR}L_{IR}$ using the above $L_{IR}/L_{1.6}$-based relation. We find that now we recover the 100\,Myr averaged SFR much better than any of the relations shown in Figure\,\ref{fig:lum_vs_sfr}. It should be emphasized however, that this good agreement for our simulations stems from the fact that we calibration our relation on these same simulations. More work is needed to test the full range of applicability of this method. Each method discussed here was calibrated on particular SFHs. It may therefore be useful to compare them as a means of finding galaxies in this rapidly quenching regimes -- these should have $k_{IR,here}\approx0$ but $k_{IR,Boquien}>0.4$. 

But how can we understand this role of $L_{IR}/L_{1.6}$ in deriving $L_{FUV,corr}$? Figure\,\ref{fig:irratio_corr} shows us that $k_{IR}$ scales linearly with $L_{\rm{IR}}/L_{1.6}$ until $L_{\rm{IR}}/L_{1.6}\sim1$ after which it is essentially zero. We believe this is driven by the role of intermediate age stars in dust heating. Such stars are less powerful than the O/B stars associated with recent star-formation and less of their light is dust obscured.  Therefore galaxies where the dust is heated by intermediate age stars are going to have a lower IR emission relative to their stellar mass compared to galaxies where the dust is primarily heated by younger stars. The `break' in the relation is reached when role of recent star-formation in dust heating becomes negligible relative to that of intermediate age stars (where sSFR is below a few $\times10^{-2}$[/Gyr]). 

We stress however that while this relation is found to work for our particular simulations, further study is needed to refine it and in the process clarify the basic physical picture discussed above.

\subsubsection{UVJ diagnostic diagrams \label{sec:uvj}}

\begin{figure}
\includegraphics[width=0.45\textwidth]{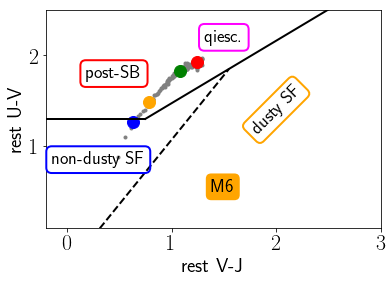}
\includegraphics[width=0.45\textwidth]{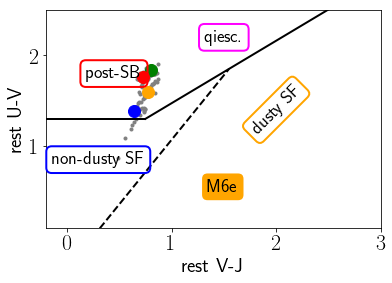}
\includegraphics[width=0.45\textwidth]{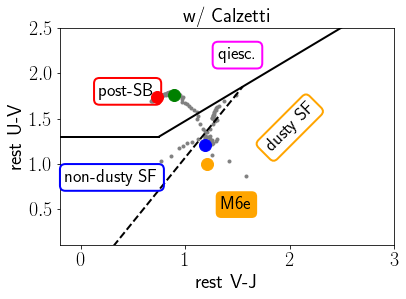}
\caption[UVJ diagrams for our simulations]{While our simulations represent dusty galaxies that at least in the early and up to (pre-)coalescence stages show significant sSFR, they nonetheless {\it never} reach the dusty star-formation locus in the diagnostic UVJ plot preferring instead the post-starburst locus. For clarity we only show one disk and one merger simulation -- they exemplify the behavior of the rest. Each point is the averaged over viewing angle. The solid lines distinguish star-forming galaxies from post-starburst and quiescent galaxies \citep{Whitaker2012,Martis2016}. The dashed line represents $A_V=1$ and separates non-dusty from dusty star forming galaxies \citep[based on][]{Martis2016}. }
\label{fig:uvj}
\end{figure}

In the previous sections, we identified that in particular, the degree of dust obscured star-formation can be over-estimated in quenching and post-starburst systems, even though in some massive, gas-rich mergers the luminosity in this quenching regime can still exceed 10$^{12}$L$_{\odot}$ i.e. a ULIRG. Indeed \citet{Sklias2017} use the rest-frame UVJ plot to find that nearly 10\% of {\sl Herschel} galaxies are likely post-starbursts (see Section\,\ref{sec:discussion}). In Figure\,\ref{fig:uvj} we show the rest-frame U-V vs. V-J diagram with three lines used to differentiate (non-)dusty star-forming, post-starburst galaxies, and quiescent galaxies \citep[e.g.][]{Whitaker2012,Martis2016}. 

We find that our simulations mostly populate the post-starburst regime regardless of their actual evolutionary stage. For simplicity in Figure\,\ref{fig:uvj} we only show two cases: for the merger M6e (\emph{bottom}) and isolated disk M6 (\emph{top}) simulations. These however, are representative of the rest of our simulations. The discrepancy between UVJ-based classification and the actual simulation properties is particularly striking for the merger M6e at $t=t_2$ where it reaches a SFR of $>$3000\,$M_{\odot}$/yr, with $A_V\approx2$ and is at that point above the SFG main sequence for $z\sim2$ galaxies \citep{Rodighiero2014} that it was designed to model \citep{Hayward2011}. Yet this system would be classified as ``quenched" on the basis of UVJ colors alone. We note that the diagnostic delineations here are based on galaxy modeling that assumes a fixed dust attenuation law (the Calzetti starburst law). As discussed in the introduction, there are indications in the literature this may not be appropriate for all dusty galaxies (indeed applying it widely to IR selected galaxies is likely wrong given that it was defined for UV-bright ones).  In the bottom panel of Figure\,\ref{fig:uvj}, we tested this by taking our un-attenuated colors and applying a dust screen with Calzetti attenuation and the measured $A_V$ values, and reproduced the standard behavior in this plot -- i.e. dusty SF galaxy colors whenever $A_V>1$. Therefore our effective dust attenuation must deviate significantly from Calzetti. In the next section we explore this further.

We caution that while our simulations do not reach the dusty SFG regime of the UVJ plot, there are plenty of observed samples that do \citep[e.g.][]{Whitaker2015,Marmol2016,Straatman2016}.  In Section~\ref{sec:discussion}, we address why we believe these observed galaxies differ from our simulations. 

\subsection{Simulated attenuation curves}

\begin{figure*}[htp]
\includegraphics[width=0.45\textwidth]{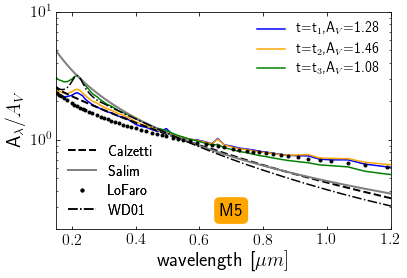}
\includegraphics[width=0.45\textwidth]{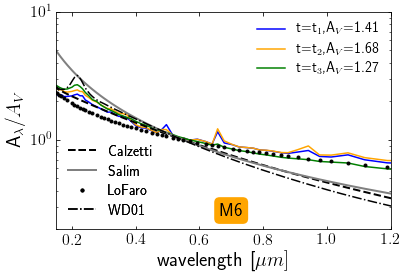}\\
\includegraphics[width=0.45\textwidth]{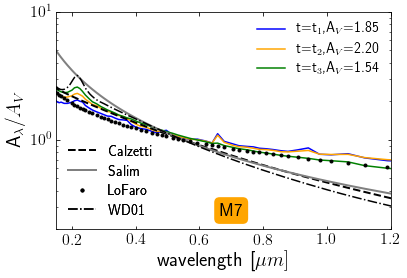}
\includegraphics[width=0.45\textwidth]{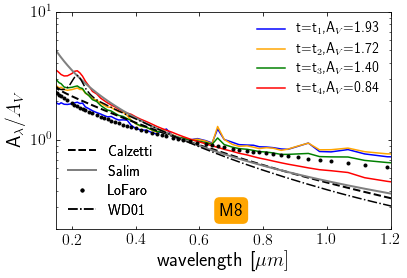}\\
\includegraphics[width=0.45\textwidth]{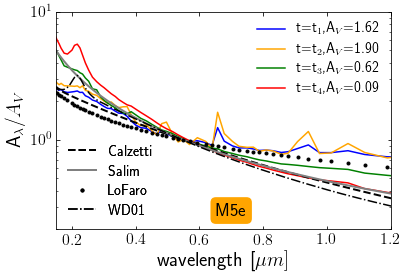}
\includegraphics[width=0.45\textwidth]{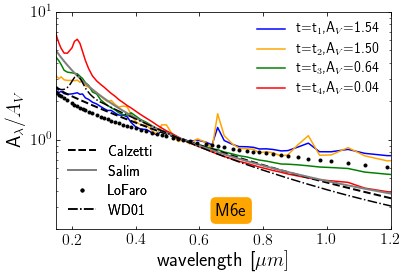}
\caption[Effective attenuation curve trends for isolated disks]{Effective attenuation curves for the isolated disk galaxies ({\it top two rows}) and the major mergers ({\it bottom row}). We show these for the representative time epochs from Figure\,\ref{fig:time_evolution}. For comparison, we overlay the \citet{Calzetti2000}, the \citet{LoFaro2017}, the \citet{Salim2018} empirical attenuation curves, as well as the MW-like extinction curve \citep{Weingartner2001}, which is the input dust model for our simulations. In the dusty epochs of each simulation we find flatter than Calzetti optical-to-NIR attenuation consistent with \citet{LoFaro2017}, while blueward of the $V$-band the attenuation curves are similar to Calzetti. In the less dusty obscured epochs (where $A_V<1$) the attenuation curves steepen considerably in both regimes. }
\label{fig:attenuation}
\end{figure*}

To estimate attenuation as a function of wavelength, we use $A_{\lambda} = -2.5 \log_{10} ( L_{\lambda} / L_{\lambda,0})$ where $L_{\lambda}$ is the post radiative transfer spectral energy distribution, and $L_{\lambda,0}$ is the pre-processed SED (see Figure\,\ref{fig:SED_summary} as an example). As discussed in Section\,\ref{sec:simulations}, the pre-processed stellar SED includes HII regions and PDRs. This means we are only discussing the attenuation done by the host galaxy at scales larger than the effective gas/star particles. We normalize these by the $A_V$ values derived for each timestep. These $A_V$ values are averaged over the 7 isotropic viewing angles in each simulation. We tested that the variation between the individual viewing angles is much smaller than the variations between timesteps seen both in the UVJ plots and the attenuation curve plots discussed here. Therefore for simplicity, we chose to focus on the variations with time and use angle-averaged values.  The thus derived attenuation curves for our fiducial epochs are shown in Figure\,\ref{fig:attenuation}. 

The effective attenuation curves for the isolated disk simulations are all consistent with attenuation curves blueward of the $V$-band ranging from Calzetti-like for the earliest $t=t_1$ epochs to typically Milky Way-like for the later epochs. Therefore we generally find steepening slopes and increasing strength of the 2175 \AA bump with decreasing sSFR (as is the trend with time for all of these isolated systems). This is consistent with observationally defined trends with sSFR \citep[e.g.][]{Kriek2013,Salim2018}). We however find grayer than Calzetti attenuation curves redward of the $V-band$ (with the exception of M8 $t_4$). Indeed, existing studies rarely address this regime, almost exclusively focusing on the short wavelength attenuation \citep[with the exceptions of][]{CharlotFall2000,LoFaro2017,Buat2018}.  Here we show the attenuation curve from \citet{LoFaro2017} who studied the attenuation up to the near-IR in IR-selected galaxies at $z\sim2$. Consistent with our results they find relatively grey attenuation in this regime. Overall the LoFaro curve is the closest to what we observe for our isolated disk simulations from the u-band through the near-IR. 

The effective attenuation curves for the merger simulations show significantly more variation than the disk attenuation curves -- likely driven by their much more variable star formation histories and geometries. But the general trends identified above continue here. Attenuation curves are fairly flat across the wavelength range (with a weak 2175 \AA bump) to start with and progressively steepen first blueward and then redward of the $V-band$. In the post-starburst stages ($t=t_4$), they show standard behavior redward of $V$ but are very steep at shorter wavelengths. 

We emphasize that in all these simulations, the input dust model is given by \citet{Weingartner2001} and therefore a pure screen extinction should follow this curve in the plots in Figure\,\ref{fig:attenuation}. The variations in the effective attenuation we observe are unrelated to any variations in the dust properties themselves -- they are entirely driven by relative star/dust geometry and the SFH of the galaxies (see trends with sSFR discussed above).  In the Discussion we use a toy model to explore further these trends in effective attenuation curves. Our conclusions, confirming earlier results \citep[see e.g.][and references therein]{LoFaro2017}, in that these trends can be understood by allowing for differential attenuation towards the younger stars vs. toward the older stars (i.e. birth cloud attenuation and diffuse ISM attenuation). SED fitting codes (e.g. MAGPHYS and CIGALE) that allows for such differential attenuation are likely to do better than ones that assume a single dust screen. 

\section{Discussion} \label{sec:discussion}
\subsection{Toy model dust attenuation}

\begin{figure*}
\includegraphics[width=0.5\textwidth]{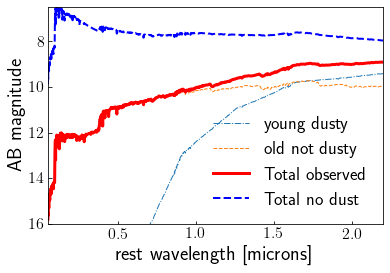}
\includegraphics[width=0.5\textwidth]{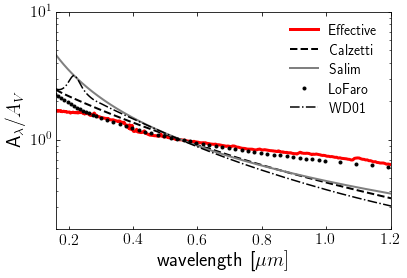}
\includegraphics[width=0.5\textwidth]{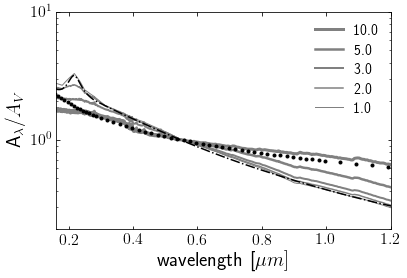} 
\includegraphics[width=0.5\textwidth]{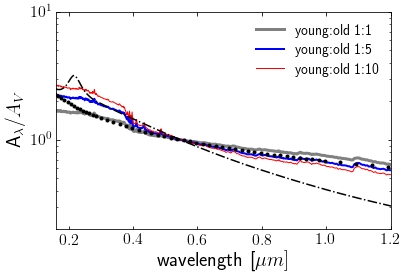} 
\caption{{\it Top-left}: An observed SED containing a dusty young population and an unobscured older stellar population (see text for details). {\it Top-right}: The effective attenuation in this case is greyer than Calzetti in the optical/near-IR regime. {\it Bottom-left}:  The effective attenuation curve in the toy model with varying $A_V$ on the young dusty component (as indicated in the legend). {\it Bottom-right}: The same for different mass ratios between the young and old components (as indicated in the legend). Here we assume a fixed $A_{V,young}=10$. In the bottom two panels, the dot-dashed curve represents a MW-like dust extinction and the black dots represent the \citet{LoFaro2017} attenuation curve.
\label{fig:eff_attenuation}}
\end{figure*}

In this section, we present our toy models that help us understand the variability in the observed attenuation curves observed in our simulations. Beforehand however it is worth reiterating some key trends we discovered in our simulations. We observed that in regime of strong star-formation as in the early stages of the isolated disks or gas-rich mergers, we have attenuation curves that are greyer than Calzetti in the near-IR regime but broadly consistent with Calzetti shorterward than the V-band (this is consistent with the findings of \citet{LoFaro2017}. Peaks in SFR that follow steadier star-formation (e.g. coalescence regime in the gas-rich mergers) lead to steepening shortward of $V$ and remain grey redward of $V$. The curves tend to steepen overall with decreasing sSFR and decreasing $A_V$. In our toy model we aim to understand all these trends. 

In Figure\,\ref{fig:eff_attenuation}, we explore the effect of combining an un-obscured older population with obscured younger population. We use the \citet{Maraston2005} models selecting two models with exponentially declining star-formation histories, where $\tau$ denotes the $e$-folding time. All models we use have solar metallicity and a Kroupa IMF. The `young' model has $\tau=0.1$\,Gyr and an age of 10\,Myrs. The `old' model has $\tau=0.5$\,Gyr and an age of 1\,Gyr. In this case, we take the young and old population to have a 1:1 ratio in stellar mass.
The young population is subject to strong screen extinction (with $A_V$=10 assuming a MW-type dust). Such $A_V$ values are not seen in whole galaxies (but see below for the effective $A_V$); however, are reasonable for the dust screen in front of actively star-forming regions alone. The old population has no dust attenuation applied to it. In this model we assume the stellar mass in the young and old component is the same. Note that in this scenario, somewhat counter-intuitively, the older population dominates the observed UV/optical, whereas the younger population dominates the near-IR. We sum the two components before and after dust extinction and compute the effective attenuation curve between them. We find attenuation greyer than the input MW extinction curve -- consistent with Calzetti in the U to V and with LoFaro in the V to near-IR regime. We note that the effective $A_V$ for this toy model is $\approx$2.5. 

In Figure\,\ref{fig:eff_attenuation} {\it bottom-left}, we show that this behavior is sensitive to the amount of attenuation toward the young component. 
The Calzetti or Milky Way laws are reproduced in the optical to near-IR regime if the attenuation toward the younger stars is lower than $A_V\approx3$ (as one would expect in UV-selected starbursts).
Significant attenuation toward the younger population coupled with significant older population present results in greyer than Calzetti curves in the near-IR although very similar to Calzetti (considering the U-V color alone) in the regime blueward of the $V-band$. Since our simulations also start off with a significant in-situ older stellar populations and we model them at epochs when they also undergo significant dust-obscured star-formation this trend is at least in qualitative agreement.

In Figure\,\ref{fig:eff_attenuation} {\it bottom-right}, we show the same model but with varying young-to-old population stellar mass ratios. This is a very rough proxy for varying the SFH in this model. We tested this by also exploring toy models with intermediate age populations. We find whenever the contribution of the young-to-old stars is below the 1:1 mass ratio than the attenuation curves are steeper than Calzetti blueward of the V-band, just as observed in our simulation -- see e.g. Figure\,\ref{fig:attenuation} M6e panel. Recall that M6e starts off with an in-situ stellar mass of $10^{11.5}M_{\odot}$ (Table\,\ref{tab:modeltable}. 

These toy models are simplistic and do not cover the full parameter space. They do however serve to exemplify how we can explain the variability in attenuation curves as a combination of differential attenuation and the specifics of the SFH.  A more detailed study of the way in which SFH or relative geometry affect the resulting attenuation curves is beyond the scope of this paper. The fact that we see comparable behavior in our simulations which have much more complexity and a range of SFHs suggests that other factors are to a large extend secondary. The core driver of this greyer than Calzetti attenuation curves toward the near-IR appears to be the combination of an older relatively unobscured stellar population and a younger heavily obscured one. Indeed, this is likely the case for many if not most dusty star-forming galaxies at cosmic noon. Our results re-enforce the conclusion that adopting a fixed Calzetti attenuation for such systems will likely result in severe errors in the derived stellar population parameters. For example \citet{LoFaro2017} show that using this more flexible attenuation curve treatment results in stellar masses on average a factor of 3 higher than using a fixed Calzetti attenuation curve (up to 10\,$\times$ for some extreme cases).  

\begin{figure*}
\includegraphics[width=0.45\textwidth]{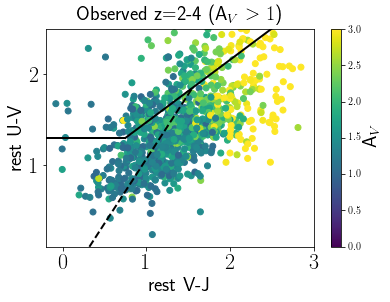}
\includegraphics[width=0.45\textwidth]{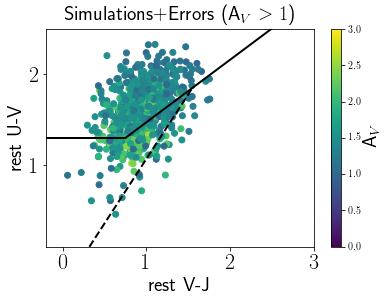} \\
\includegraphics[width=0.45\textwidth]{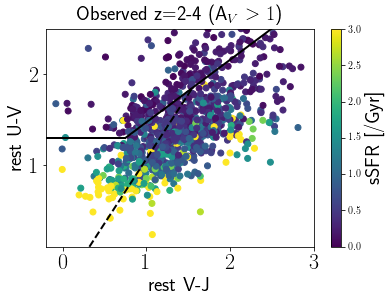}
\includegraphics[width=0.45\textwidth]{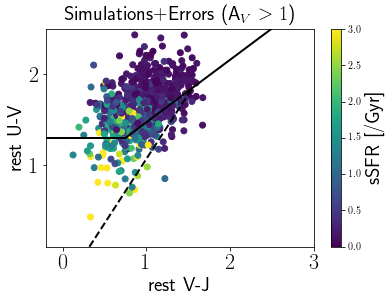}
\caption{Comparing the rest-frame UVJ colors distribution of our simulations to those of observed $z\sim2-4$ galaxies, color-coded by $A_V$ and sSFR shows us that our simulations are neither as dusty (high $A_V$) for nor as `bursty' (high sSFR) as the observed sample. These observed galaxies show that the dusty star-forming galaxies regime in the UVJ plot is dominated by extreme $A_V>2$ galaxies (these are typically found at $z\sim2-3$), while the most strongly starbursting dusty galaxies (sSFR$\gtrsim$3 Gyr; typically at $z>3$ in this sample) are found in the non-dusty star-forming galaxies locus, pushing toward the post-starburst and quiescent regimes as the sSFR drops. Our simulations agree with these trends in the regimes where these parameters overlap. 
{\it Left panels:} The UVJ colors of UltraVISTA galaxies from Martis et al. 2019 (in press). {\it 
Right panel:} The UVJ colors of all our simulations where we have $A_V>1$. We also added Gaussian errors of 0.2 to both colors. Note that we reproduce the trend with sSFR already noted in Martis et al. Our simulated colors overlap with the observed distribution modulo the fact that we do not reach as high sSFR as in our simulated library as the real galaxies in this sample.}
\label{fig:compare_martis2018}
\end{figure*}

\subsection{Comparison with observations}
\subsubsection{IR-luminous post-starburst galaxies?}
While earlier studies have already pointed out the role of intermediate age stars in dust heating \citep{Hayward2014}, this has generally been thought of as an issue for lower IR luminosities (typically $L_{\rm{IR}}<10^{10}$L$_{\odot}$). In our simulations we find cases of gas-rich massive galaxy mergers where $L_{\rm{IR}}>10^{12}$L$_{\odot}$ (i.e. ULIRGs) in a regime where the instantaneous SFR is negligible (see in particular the mergers M5e and M6e). Do such systems exist in reality and how much of an issue are they for the broader population of dusty IR-luminous galaxies at cosmic noon? Using a UV through far-IR SED fitting analysis of a large sample of  GOODS-Herschel $z\sim2$ galaxies, \citet{Sklias2017} 
find $\sim$7\% of the total sample are consistent with such rapidly declining SFHs where the true SFR is 
an order of magnitude or more below that inferred from the Kennicutt relations even using SFR$_{IR+UV}$. This just under 10\% of the {\sl Herschel} galaxies estimate is a small but non-negligible part of the whole population. It is in line with conclusions that the typical dusty IR-luminous galaxy at cosmic noon is probably not associated with a major merger (in our simulations such rapid quenching is associated with major mergers) but rather the normal mode star formation in largely isolated disks \citep[e.g.][]{Rodighiero2011}. This of course, coupled with a steep stellar mass function, explains the relatively low fraction of {\sl Herschel} ULIRGs that are post-starburts. Using the difference between the SFR derived using say the Boquien et al. method and our proposed $L_{\rm{IR}}/L_{1.6}$ method can help identify such systems relatively cheaply in large samples. Seeing their fraction in different mass bins would be a means of identifying the role of rapidly quenching systems (potentially post-coalescence mergers) as a function of mass. This requires cosmic noon samples that reach below the confusion-limited {\sl Herschel}-samples, for example the upcoming LMT Toltec surveys\footnote{For more details see \url{http://toltec.astro.umass.edu/}}. 

\subsubsection{UVJ trends -- observed vs. simulated}
The {\sl Herschel} $z\sim2$ sample of \citet{Sklias2017}, the 24\um-selected $z\sim2$ galaxies of \citet{LoFaro2017}, and many other studies all do find that dusty IR-luminous systems at $z\sim2$ do largely populate the dusty SFG part of the UVJ diagram. This is despite indications of effects of varaible SFHs and variable dust attenuation (recall that \citet{LoFaro2017} find greyer than Calzetti attenuation in the V-J regime). So why do our simulations lack {\it any} galaxies in the dusty star-forming galaxies part of the UVJ plot? 

To help us understand this behavior, we want to compare our simuations more carefully with the recent paper Martis et al. (2019, in press). This analysis presents a sample of $z>1$, $\log(M_*/{\rm M_{\odot}})>10.5$, and $A_V>1$ galaxies drawn from the COSMOS UltraVISTA catalog (parameters based on MAGPHYS SED fits). This is a good comparison sample as all our simulations are high mass ($\log(M_*/{\rm M_{\odot}})>10.5$) and here we will focus only on the timesteps in our simulations library where $A_V>1$ -- to approximate the Martis et al. selection. The $A_V$ for this observed sample is the MAGPHYS output of this quantity. It is the effective attenuation at the V-band computed by comparing the SEDs before any dust attenuation and after (including both birthcloud and diffuse ISM attenuation).

Martis et al. explore the behavior of their galaxies in the UVJ plot and indeed find some dusty star-forming galaxies that do not lie where expected in this plot due to differential attenuation (the difference between birthcloud and ISM attenuation). In Figure\,\ref{fig:compare_martis2018} we compare the UVJ plot distribution of their galaxies and our simulations color-coded both by $A_V$ and by sSFR. 

In the top panels we highlight the trend with $A_V$. Relative to the nominal line differentiating dusty star-forming galaxies from non-dusty star-forming galaxies and quiescent galaxies, the whole distribution for the observed galaxies is shifted toward bluer V-J colors consistent with greyer effective attenuation in the V-J regime. While there have plenty of galaxies in the dusty star-forming range, they are predominantly $A_V>2$ not $A_V>1$. Our simulations rarely reach that high values of $A_V$, typically being being $A_V\approx1-2$.  The lack of very high $A_V$'s in our simulations may be explained by the fact that for star particles with age $<10$\,Myr, the input stellar SED includes both the Starburst99 model and its processing through MAPPINGSIII which includes birthcloud attenuation. This additional attenuation is not accounted for in our analysis where we derive $A_V$ based on the ratio of the SEDs pre- and post-radiative transfer, see Figure\,\ref{fig:SED_summary}. On the other hand, our simulations are far from the dust screen model for host galaxy attenuation as in MAGPHYS for example. It is worth noting that even without resolving the stellar birthclouds, they are composed of SPH gas/dust particles which are concentrated in areas of enhanced star-formation activity. Differential attenuation is therefore still taking place thus accounting for the greying of the effective attenuation curve \citep[an effect first described in][]{WittGordon1996}. Therefore this comparison is still useful is telling us that the qualitative behavior of the observed galaxies sample is consistent with our results -- with bluer than expected V-J colors suggesting greyer than Calzetti attenuation curves in this range.

In the bottom panels of Figure\,\ref{fig:compare_martis2018}, we plot the same data now color-coded by sSFR. 
We find a general trend of incresing sSFR toward bluer colors and vice versa. This is again as expected from differential attenuation which resulting in greying of the effective dust attenuation curves.  In particular, we find a population of very blue dusty galaxies (predominantly in the range $z=3-4$) that fall in the non-dusty star-forming galaxies regime.  This trend among the observed galaxies is reproduced by the simulated galaxies modulo differences in the relative frequencies in different parts of the diagram due to selection effects -- our particular simulations library is biased toward somewhat less extremely dusty and lower sSFR systems than the observed sample. 
These trends need to be investigated using the more representative populations of cosmic noon galaxies in cosmological simulations -- this investigation however is beyond the scope of this paper. For example, \citet{Narayanan2018} looked at the variability in the dust attenuation curve in Illustris galaxies and found largely Calzetti-like curves by $z\sim6$ (consistent with relatively low dust obscuration and high sSFR), and show a large variability at lower redshifts (especially $z<3$) (consistent the combined effects of star-formation history and differential attenuation as explored in our toy model). They however only focused on the short wavelength part of the attenuation curve and did not explore its behavior out to the near-IR which is critical for the UVJ plot. 

\subsection{Caveats}
The analysis of this paper is based on simulations and therefore the conclusions herein are contingent on the accuracy of the assumptions embedded in our simulations \citep[see e.g.][and references therein]{Popping2017}. These are particularly uncertain when it comes to the multiphase ISM geometry and making more realistic simulations in this regard is needed. As discussed in Section\,\ref{sec:simulations}, we find the trends to be the same between the {\sc mp-on} and {\sc mp-off} treatments. We do not however have this variation for our entire library precluding a more in-depth analysis. These variations are fundamentally only crude approximations to the real ISM of galaxies. 

Our results are based on idealized simulations where the galaxies cosmological environment is neglected. By contrast, \citet{Narayanan2018} use a combination of zoom-in simulations (similar to this work) as well as a 25\,Mpc$^3$ box cosmological simulation to examine in particular the variability in UV slope and 2175\AA\ bump strength. This level analysis but extending to the near-IR and explicitly addressing the UVJ colors of galaxies is needed. As shown above, our particular simulation library is neither as dusty nor as `bursty' (i.e. high sSFR) as typical observed samples of dusty massive galaxies at cosmic noon.

In this paper we also do not explore the effects of different dust composition -- we assume MW-like dust throughout. This helps us highlight the variability in the attenuation curve that is due to factors independent of the actual dust properties. However, in \citet{Safarzadeh2017}, these are shown to be non-negligible when it comes for example to the observed IRX values.  Our analysis therefore highlights how differential attenuation and different star-formation histories can lead to significant observed variability of the attenuation curve (and hence UVJ colors) even for the same dust model.    

\section{Summary \& Conclusions} \label{sec:conclusions}

It is common to consider that a direct measure of dust emission (i.e. total IR luminosity) is the best proxy for the dust obscured star-formation, especially for IR luminous systems. However, direct detection of dust emission is strongly limited by the confusion of existing far-IR instruments, so we rely on its indirect assessment from constraints on dust attenuation in shorter wavelength data. For example, the role of dust obscured star-formation in larger galaxy samples as a function of stellar mass and redshift is assessed using the UVJ diagnostic plots  \citep[e.g.][]{Whitaker2012,Straatman2016,Martis2016}, where galaxies are assumed to follow the same Calzetti dust attenuation curve. In this paper, we use hydrodynamic simulations with radiative transfer to examine how well can we recover the simulated dust obscured star-formation using both the direct dust emission and the indirect UVJ plot.  Our conclusions are as follows:

\begin{itemize}
    \item The total IR luminosity is not a good proxy for SFR in regimes when the galaxies are rapidly quenching or post-starburst. This regime can reach $L_{IR}\sim10^{12}$L$_{\odot}$ for extreme systems. IR-luminosity-based methods of estimating the role of dust obscured star-formation in a given population are an overestimate due to the role of intermediate-age stars in dust heating.
    \item Composite FUV+IR relations \citep[as in][]{Boquien2016} do work much better at recovering the true SFR, but again are an overestimate in rapidly quenching regimes as expected post-coalescence in major mergers. We find this can be corrected by using the $L_{IR}/L_{1.6}$ ratio. 
    \item Our simulations fail to populate the dusty star-forming galaxy regime of the UVJ plot. This is due to variable attenuation curves that in particular are much grayer in the optical-to-nearIR regime than the Calzetti curve \citep[in agreement with][]{LoFaro2017}.  
    \item A comparison with observed samples suggests the galaxies that do populate the dusty SFG part of the UVJ plot are likely a particular sub-set of all dusty star-forming galaxies -- one that is not represented in our limited simulations library. Therefore, UVJ-based methods of estimating the role of dust obscured star-formation in a particular population are an underestimate due to the effects of variable dust attenuation curves.
\end{itemize}

\acknowledgements
We thank V\'eronique Buat for a careful reading of and useful feedback on the paper pre-submission.  We also thank Desika Narayanan and Samir Salim for useful discussions. 

\bibliography{roebuck_ppii_arxiv}

\end{document}